# Multiclass Sentiment Prediction for Stock Trading

Marshall McCraw
*Department of Mechanical and Aerospace Engineering*
*George Washington University*
Washington, D.C.
mmccraw98@gwmail.gwu.edu

*Abstract*—Python was used to download and format NewsAPI article data relating to 400 publicly traded, low cap. Biotech companies. Crowd-sourcing was used to label a subset of this data to then train and evaluate a variety of models to classify the public sentiment of each company. The best performing models were then used to show that trading entirely off public sentiment could provide market beating returns.

## I. INTRODUCTION

Experience shows that the price of an asset on a stock market is rarely dependent on its fundamental financial value. Many times, 'animal spirits' may significantly alter an asset's price due to some general public sentiment which may cause traditional investment methods to be unintentionally affected by sensationalist driven trends. Therefore, if the public sentiment about a specific company were able to be quantified and tracked, an investor should be able to use this signal for productive trading as effectively as any other investing tool.

## II. DATA SOURCING

### A. Web Scraping

In order to quantify the public sentiment expressed about a specific company, news article text data were downloaded and used to train a series of Sci-Kit Learn classifiers to accurately predict the conveyed sentiment of a news article. As text related classification problems generally have high dimensional feature space, their classifiers tend to suffer from the 'curse of dimensionality' and may overfit the data without a suitably large training dataset [1]. This highlights the need for a reliable data source that has exceedingly high to no limit on data download. Internet based news articles are programmatically accessible through a wide variety of techniques; the first attempt at sourcing data used custom scripts known as webscrapers, the second and final attempt used a third party API which allowed for much simpler data sourcing.

In total, two webscrapers were built and configured for use on Google News and, the crowd-sourced financial news platform, SeekingAlpha. A list of 400 publicly traded, small cap. Biotechnology companies was created with the YahooFinance stock screening tool. The company names and stock tickers on this list were used to automatically create search queries for each company which were then used to provide a list of urls over which each webscraper would operate.

The first webscraper tested used the requests library to quickly loop through and download the html source from each url on the list. While able to offer suitably high download rates, this script would quickly get flagged and subsequently blocked with captchas. To avoid this, the browser automation library, Selenium, was used to mimic more human behavior through a series of randomized interactions with features on each site. While navigating the webpages, the Selenium webscraper would spend different amounts of time on each page before downloading the html source and using keyboard inputs to move on to subsequent pages. While less systematically detectable, this approach would inevitably get captcha blocked and was too slow to be used for practical large scale data collection, though its ease of reconfiguration for use on different websites makes it a focus of future work.

The final method of data sourcing used was the NewsAPI, a program which, among other things, compiles every news article from over 50,000 news sources to allow query based programmatic searches. Further, NewsAPI allows developers up to 500 daily free searches that can return 100 articles within a time period 15 minutes to one month before the search is made. While the one month time limit restricts the scope of the available data, the fast and easy to use search protocol rendered the NewsAPI as an attractive solution.

### B. NewsAPI

The same query generation script was used to direct the NewsAPI search for each company on the screened list. For each article in the search, the NewsAPI returns the title, description, content snippet, author, and publication date. Each of these features is then saved to a .csv once the search is complete. Due to the limited search volume and period, this program was run daily to collect as much article data for each company as possible. After two weeks of daily data collection, each company's .csv files were consolidated, keeping only the unique entries, and formatted to utf-8 encoding for the sake of commonality. Additionally, the combination of each article title, description, and content was added as a new feature called 'combination'. After completing this process, the four features (title, description, content, and combination) were copied into separate .csv thereby generating four separate datasets. Once labeled, these datasets would serve as the data on which sentiment classification models would be trained, tested, and evaluated.

*C. Crowdsourcing*

As the data downloaded from the News API was unlabeled, Amazon Mechanical Turk (AMT), a crowd-sourcing platform, was used to create labels for the data so that models could be trained. AMT connects requesters, who have tasks which can only be completed with human intelligence, with workers who will complete such Human Intelligence Tasks (HITs) for payment. AMT gives researchers the ability to rapidly create labels for previously unlabeled datasets by using the large number of workers available on the site.

200 entries from each of the four datasets were randomly sampled and uploaded to AMT to be labeled. For each entry, the HIT prompted the workers to label the sentiment conveyed through the given body of text as it related to a specified company. An example of such a HIT is available in Figure 1.

*Figure 1: Sample Title Classification HIT*

Each HIT asked workers to specify whether the text conveys a positive, neutral, or negative sentiment about the company. The given answer would then be stored along with the workers identification and work time. A total of three unique workers would supply answers for each HIT so that the 'correct' answer could be derived from the median of the three labels. Results from the HITs are not without flaws, for instance, a common tactic for workers looking to game the system is to click through HITs as fast as possible without providing an accurate answer. If caught, AMT enables requesters to easily reject inaccurate responses and work done by cheaters. Though skilled cheaters may provide reasonable looking answers to evade cursory detection, most can be identified by filtering workers by average answer time and accuracy on a set of pre-labeled HITs often referred to as Gold HITs. Figure 2 demonstrates there were several trends amongst the average work time in each dataset. There appear to be two clear distributions of workers who averaged less than 15 seconds per HIT with one of these distributions centering about the 5 second mark. As the mean work time across all workers on all datasets was 86 seconds, it is highly unlikely that workers who fall in the sub 15 second distributions are providing meaningful answers.

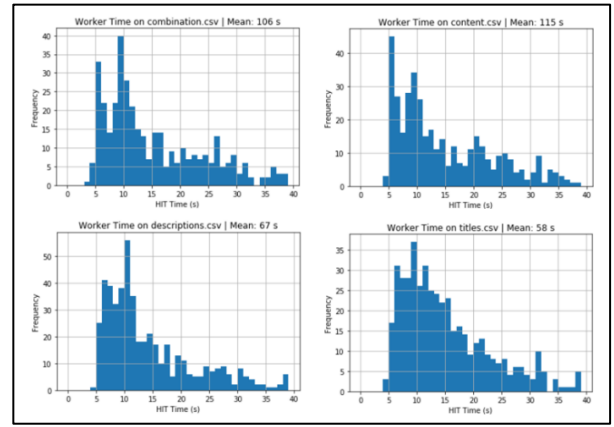

*Figure 2: Histograms of Worker Time on Each Dataset*

To more rigorously determine which data were falsified, a cheater screener was made which identified the Worker IDs of those who had an accuracy on the Gold HITs and average work time less than 30% of the average values for all workers. This screener identified over 10 suspected cheaters whose work was then more closely examined. Plotting a histogram of each suspected cheaters answers would provide the final determination on whether a given worker was cheating. In order to minimize unfair rejections, cheaters were only rejected if they were undoubtably cheating. Often, suspected cheaters were workers who had only answered one or two HITs and therefore, could not meaningfully be marked as cheating. The most flagrant types of cheaters were workers who repeatedly entered the same answer and workers who randomly entered answers getting few correct.

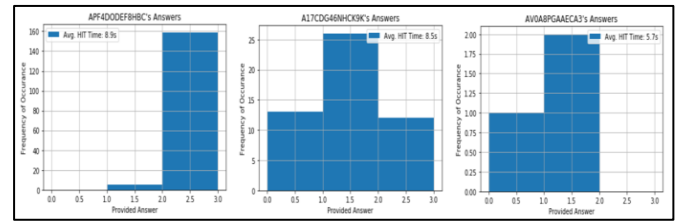

*Figure 3: (a) Unskilled Cheat, (b) Skilled Cheat, (c) Indeterminate*

In total, 6 workers had their work rejected due to the large number of inaccurate answers provided and the rapid rate at which answers were supplied. Following rejection, AMT relists each HIT to be labeled by new workers. Once the final set of data returned, the distribution of answers was compared to the distribution of Gold HIT labels for a general comparison and sanity check seen in Figure 4. While there were minor differences, the general trends were relatively consistent with a bias towards positive and neutral sentiment. The only major difference was in the content dataset; as the content dataset Gold HIT distribution stuck out from the other AMT and Gold labeled sets, it is possible that the content dataset Gold HITs were inaccurate themselves and this discrepancy was ignored.

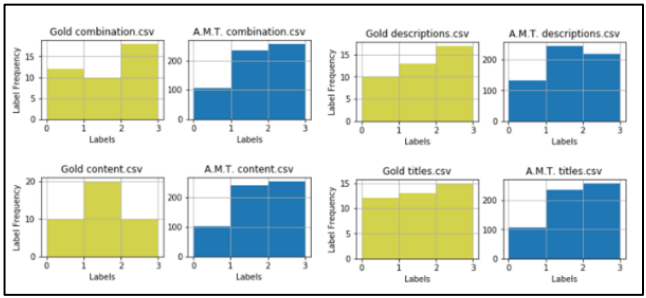

*Figure 4: Distributions of Gold HITs and AMT Results*

To further verify the validity of the crowdsourced data, Fleiss' Kappa, a measure of answer agreeance between workers, was calculated for each dataset [2]. Bound by 1 in the positive domain, a high value of Kappa indicates that the labeled supplied from the AMT workers is quite reliable, whereas values below 0 suggest there is little reliability.

| Dataset | K |
|---|---|
| combination | -0.019 |
| content | -0.020 |
| description | -0.037 |
| titles | -0.024 |

*Figure 5: Fleiss' Kappa by Dataset*

The resulting score for each dataset showed little to no agreeance between workers indicating that there is little consistent agreement amongst the data. This lack of consistency should be expected for many reasons: there is very little objectivity involved in sentiment determination as it relies on the interpretation of a reader, there are three categories which allows for more answer variance, and the large number of neutrally rated text may suggest some workers interpreted less polarizing text as neutral while others may have interpreted the same text as polar. While the results from the AMT HITs were scattered and inconsistent, the median worker label was taken for each sample to provide a finalized group of labels for each dataset.

III. MODEL SELECTION, EVALUATION, AND TRAINING

Before this data could be used for model training, a series of preprocessing operations were performed.

*A. Data Preprocessing*

TF-IDF: To make the text data computer interpretable, Term Frequency - Inverse Document Frequency (TF-IDF) transformations were fit and subsequently applied to each dataset. TF-IDF transformations, initially vectorize the text data by tokens (unique words or phrases) and then weights each token proportionally to their frequency of occurrence within the text while also decreasing the weight of tokens with the highest frequencies - thereby reducing the significance of words like 'the' which provide little contextual meaning [3].

SVD: A TF-IDF vectorized body of text will generate a dimensionally large feature space as each unique token is encoded to its own dimension. As mentioned earlier, certain words like 'the' do little in differentiating text by meaning and may lead to overfitting. A common scheme to avoid such issues is to drop redundant features from the data through a process known as dimensionality reduction. The method of dimensionality reduction used was Truncated Singular Value Decomposition (SVD). SVD determines which features have the highest covariance and, through eigen decomposition, keeps the top N variant features – thereby dropping redundant terms and reducing the dimensionality of the feature space [4].

*B. Initial Model Selection*

Two TF-IDF vectorizers were fit to each dataset; one which tokenized all unigrams and bigrams whereas the second tokenized only unigrams. Following vectorization, the data were then used to perform grid search cross validation on a series of model classes, thereby tuning the hyperparameters of each model to provide the most accurate performance for each class and dataset. The four model classes evaluated for each dataset were Logistic Regression (LR), Multinomial Naïve Bayes (MNB), Radial Basis Function Support Vector Machine (RBF-SVM), and K-Means Clustering (KM). These model classes were chosen for showing experiential success on similar text-based multi-class classification problems. The model tuning process was then repeated on vectorized data which had been reduced to 100 dimensions through SVD. The best model hyperparameters for each model class and dataset are attached in Figure A6.

*C. Model Evaluation*

The final models for each dataset were chosen through an additional round of 10-fold cross validation and separated by a specific metric. Since the models in this problem are intended to provide investment signals, the identifying metric should be chosen to minimize the risk of loosing money. If using these models for a long only investment strategy, the only risk of incurring loss occurs with false positive predictions while false negative predictions only cause an investor to lose out on the potential to earn money. Generally, false positives should be minimized as much as possible while false negatives are permissible, but not ideal. Therefore, the best performing model would provide as few false positives and a small number of false negatives. Following these guidelines, the metric by which the best performing models should be determined was selected to be the True Positive Rate, otherwise known as Recall (Equation 1).

$$R = \frac{True\ Positives}{True\ Positives + False\ Positives} \qquad (1)$$

As previously noted, each tuned model was then evaluated through 10-fold cross validation on the labeled training data. Once this process was complete, an average mulit-class Recall value for each model was reported. The model with the highest Recall values for the positive and negative classes was selected for each dataset and highlighted in Figure A7. The last step in model selection compared the Recall scores for the previously noted 'best models' to the Recall scores of these models when the training data were equally weighted. This process used 3-fold cross validation where each fold had an equal distribution of classes. The model for the title dataset was the only one which benefitted from equal class weighting. After determining the best performing model for each dataset, the model parameters and preprocessing operations were saved and listed in Figure 8.

| Dataset | Model Class | Preprocessing | Hyperparameters | Negative Recall | Neutral Recall | Positive Recall |
|---|---|---|---|---|---|---|
| Combo | K-Means | TF-IDF Bigrams | n: 3 | 0.46 | 0.34 | 0.17 |
| Content | K-Means | TF-IDF Bigrams + SVD | n: 3 | 0.32 | 0.29 | 0.40 |
| Description | RBF SVM | TF-IDF Bigrams + SVD | C: 100 γ: 0.01 | 0.53 | 0.54 | 0.40 |
| Title | K-Means | TF-IDF Bigrams + Equal Weight | n: 3 | 0.71 | 0.23 | 0.17 |

Figure 8: Final Models and 10-Fold CV Results

### D. Final Model Training

After tuning and evaluating, the best fitting model for each dataset was then retrained on the entirety of the training data and pickled along with any relevant preprocessing transforms to a .model file. For predictive use, the .model file for each dataset can be loaded and new data can be passed into the predict function to create a sentiment classification.

## IV. APPLICATION AND CONCLUSIONS

### A. Trading Simulation

Finally, to test the practicality of trading based off predicted sentiment values, all the article data collected over the past few weeks were used to develop a rudimentary trading simulation where assets would be purchased and sold based off of sentiment predictions. Historical closing price data for each asset was acquired with the Pandas-Datareader library so that each asset could be realistically simulated from March 9 to April 7. Each asset was assumed to have an equal initial investment amount, therefore only percent return on investment was necessary to track model performance

$$ROI = \frac{Final\ Closing\ Price - Initial\ Closing\ Price}{Initial\ Closing\ Price} \quad (2)$$

Articles published on each simulated day were preprocessed and fed into the set of models for each company. The models output generated a set of sentiment predictions; for simplicity and stability, the mean of these four predictions was recorded and used as the primary trading signal.

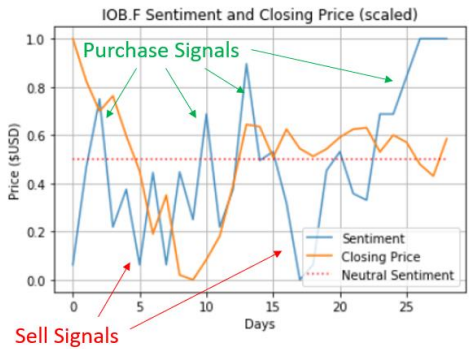

Figure 9: Scaled Price and Sentiment Predictions for $IOB.F
(sentiment predictions lead price, buy when scaled sentiment > 0.5, sell when scaled sentiment ≤0.5)

If the mean sentiment was greater than one, then the algorithm purchased the asset in question, if it was less than one and the asset was owned, then it was sold. The percent return on investment was calculated for each asset at the end of the simulation period.

### B. Results

The results where dependent upon the number of total articles per company – where, the best test results came from assets with more than 150 total articles. Further increasing the required number of articles for each asset only served to decrease the number of assets in the simulation and introduce greater variability into the results. In the case of a 200 article minimum per company, the algorithm never profited, as seen in the simulation results in Figure 8. On a similar note, including assets with very few articles also adds variability and makes the algorithm vulnerable to unpredictable price movements since there are assets which have days without new articles.

| Min. Articles | Avg. ROI | Max. Win | Max. Loss | Avg. Win | Avg. Loss | W/L Ratio |
|---|---|---|---|---|---|---|
| 200 | -5.80% | 11% | -50% | 7.0% | -23% | 1.3 |
| 150 | 3.13% | 79% | -50% | 14% | -13% | 1.6 |
| 100 | 1.50% | 79% | -50% | 13% | -15% | 1.4 |
| 50 | 0.27% | 79% | -50% | 12% | -13% | 1.1 |
| 0 | 0.97% | 100% | -50% | 13% | -11% | 1.0 |

Figure 10: Statistics from Trading Simulation

An unrestricted algorithm tends to have diminished returns as average earnings and average losses are each diluted by the large number of assets held. It is important to note that this scenario mimics the traditional investing paradigm of risk avoidance through portfolio diversity. While this generates smaller returns, the potential downside is minimized providing much smaller variance within percent lost and a lower risk investing scheme.

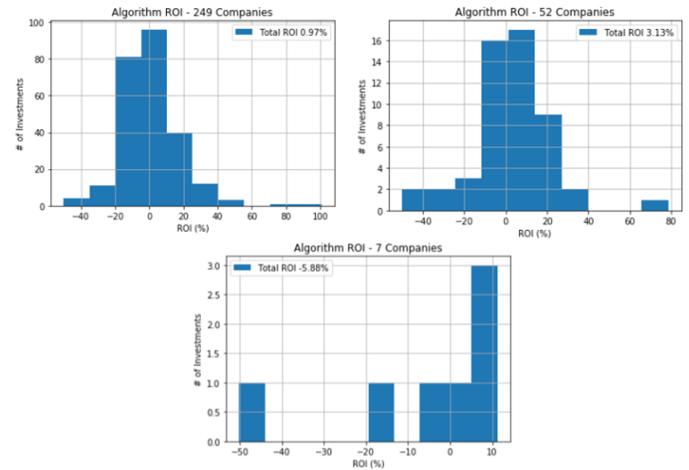

Figure 11: ROI Distributions (too few assets varies results, too many assets dilutes potential)

Greater returns are achievable by requiring enough articles to reduce the amount of undetectable price movements while also maintaining a large number of purchasable companies to reduce variability in returns. The data seem to suggest that such a scheme provides the greatest ratio of returns to losses while still maintaining a relatively low percent loss and generating 3.1% ROI. Potentially more notable, this algorithm outperforms the benchmark of the Nasdaq, S&P 500, as well as the Dow Jones which generated -3.2%, -5.0%, and -0.08% return on investment throughout the same testing period.

## C. Conclusions

While the models reported on were able to guide a basic trading algorithm to outperform the markets for an entire month, more rigorous back testing is needed in the simulation in order to generate meaningful data. One month is simply not enough time to get a full characterization of the models' behavior, especially due to the unstable and abnormal market conditions that have been induced by the COVID-19 pandemic. This said, the results, though temporally limited, are highly promising and provide an impressive foundation for future work. The next planned iteration of the trading algorithm will have three areas of improvement: Data Sourcing, Data Labeling, and Algorithm Complexity.

A 'smarter' version of the Selenium powered webscraper which could operate on numerous financial news sources could download far more relevant article data than the NewsAPI. This webscraper would simply switch to a different news source when a captcha was triggered and return once the captcha cool down period expired. This webscraper would provide a more continuous, albeit smaller stream of more relevant data. Alternatively, if the NewsAPI is to continue to be used, a better query generation script would increase article relevancy thereby providing clearer samples for the models as well as cause less confusion within the AMT workers during the crowdsourced labeling process. Along the same line, having a larger labeled dataset provided by additional crowdsourcing would provide better performing models (perhaps the additional cost could be covered by any potential earnings from this algorithm). The trading algorithm itself could be improved by taking into consideration all the predicted sentiment values for each feature instead of an average. As the predicted sentiment seen in Figure 8 was quite sharp, potentially smoothing the sentiment signals would yield better returns. Finally, the algorithm could benefit from the inclusion of other fundamental financial metrics to hedge against irrational model predictions.

## D. Figures and Tables

| Dataset | Model Class | Model Hyperparameters | | | |
| --- | --- | --- | --- | --- | --- |
| | | Features 1 | Features 2 | SVD Features 1 | SVD Features 2 |
| Combo | Log. Reg. | C: 1e-4 | C: 1e-4 | C: 1e-4 | C: 1e-4 |
| | Multinm. Nve. Bys. | α: 100 | α: 100 | α: 100 | α: 100 |
| | RBF SVM | C: 1e-5 γ: 1e-8 | C: 1e-5 γ: 1e-8 | C: 1e-5 γ: 1e-8 | C: 1e-5 γ: 1e-8 |
| | K-Means | n: 3 | n: 3 | n: 3 | n: 4 |
| Content | Log. Reg. | C: 1e-4 | C: 1e-4 | C: 1e-4 | C: 1e-4 |
| | Multinm. Nve. Bys. | α: 100 | α: 100 | α: 100 | α: 100 |
| | RBF SVM | C: 1e-5 γ: 1e-8 | C: 1e-5 γ: 1e-8 | C: 1e-5 γ: 1e-8 | C: 1e-5 γ: 1e-8 |
| | K-Means | n: 3 | n: 3 | n: 3 | n: 3 |
| Description | Log. Reg. | C: 1 | C: 10 | C: 1 | C: 10 |
| | Multinm. Nve. Bys. | α: 100 | α: 1 | α: 100 | α: 1 |
| | RBF SVM | C: 100 γ: 1 | C: 100 γ: 0.01 | C: 100 γ: 1 | **C: 100 γ: 0.01** |
| | K-Means | n: 3 | n: 3 | n: 3 | n: 3 |
| Title | Log. Reg. | C: 0.1 | C: 0.1 | C: 0.1 | C: 0.1 |
| | Multinm. Nve. Bys. | α: 100 | α: 100 | α: 100 | α: 100 |
| | RBF SVM | C: 1 γ: 10 | C: 1 γ: 1 | C: 1 γ: 10 | C: 1 γ: 1 |
| | K-Means | n: 3 | n: 3 | n: 3 | n: 3 |

*Figure A6: Model Hyperparameters from Grid Search CV*

| Dataset | Model Class | Model Recall by Class (Negative, Neutral, Positive) | | | |
| --- | --- | --- | --- | --- | --- |
| | | Features 1 | Features 2 | SVD Features 1 | SVD Features 2 |
| Combo | Log. Reg. | 0.00, 1.00, 0.00 | 0.00, 1.00, 0.00 | 0.00, 1.00, 0.00 | 0.00, 1.00, 0.00 |
| | Multinm. Nve. Bys. | 0.00, 1.00, 0.00 | 0.00, 1.00, 0.00 | 0.00, 1.00, 0.00 | 0.00, 1.00, 0.00 |
| | RBF SVM | 0.00, 1.00, 0.00 | 0.00, 1.00, 0.00 | 0.00, 1.00, 0.00 | 0.00, 1.00, 0.00 |
| | K-Means | 0.16, 0.59, 0.27 | **0.46, 0.34, 0.17** | 0.14, 0.54, 0.30 | 0.05, 0.22, 0.06 |
| Content | Log. Reg. | 0.00, 1.00, 0.00 | 0.00, 1.00, 0.00 | 0.00, 1.00, 0.00 | 0.00, 1.00, 0.00 |
| | Multinm. Nve. Bys. | 0.00, 1.00, 0.00 | 0.00, 1.00, 0.00 | 0.00, 1.00, 0.00 | 0.00, 1.00, 0.00 |
| | RBF SVM | 0.00, 1.00, 0.00 | 0.00, 1.00, 0.00 | 0.00, 1.00, 0.00 | 0.00, 1.00, 0.00 |
| | K-Means | 0.07, 0.56, 0.15 | 0.22, 0.46, 0.29 | 0.23, 0.42, 0.26 | **0.32, 0.29, 0.40** |
| Description | Log. Reg. | 0.18, 0.77, 0.42 | 0.27, 0.73, 0.42 | 0.18, 0.63, 0.45 | 0.39, 0.48, 0.44 |
| | Multinm. Nve. Bys. | 0.00, 1.00, 0.00 | 0.26, 0.74, 0.48 | 0.00, 1.00, 0.00 | 0.00, 1.00, 0.00 |
| | RBF SVM | 0.27, 0.77, 0.4 | 0.27, 0.73, 0.41 | 0.33, 0.56, 0.44 | **0.53, 0.54, 0.4** |
| | K-Means | 0.39, 0.2, 0.47 | 0.56, 0.24, 0.21 | 0.32, 0.32, 0.32 | 0.25, 0.57, 0.22 |
| Title | Log. Reg. | 0.00, 0.92, 0.15 | 0.00, 0.95, 0.13 | 0.00, 0.82, 0.32 | 0.00, 0.84, 0.26 |
| | Multinm. Nve. Bys. | 0.00, 1.00, 0.00 | 0.00, 1.00, 0.00 | 0.00, 1.00, 0.00 | 0.00, 1.00, 0.00 |
| | RBF SVM | 0.23, 0.84, 0.22 | 0.23, 0.82, 0.24 | 0.20, 0.82, 0.23 | 0.20, 0.59, 0.47 |
| | K-Means | 0.61, 0.17, 0.19 | 0.52, 0.37, 0.12 | 0.26, 0.39, 0.3 | 0.22, 0.24, 0.49 |

*Figure A7: Model Recall by Class (from 10-fold CV)*

### ACKNOWLEDGMENT

Special thanks to Dr. Broniatowski and Dr. Smith for their assistance and instruction throughout the semester. This class has proven to be very useful and highly insightful.